\documentclass[conference]{IEEEtran}
\IEEEoverridecommandlockouts
\usepackage{balance}
\usepackage{ifpdf}
\usepackage{cite}
\ifCLASSINFOpdf
   \usepackage[pdftex]{graphicx}
\else
\fi
\usepackage{amsmath}
\usepackage{algorithmic}

\usepackage{array}

\ifCLASSOPTIONcompsoc
  \usepackage[caption=false,font=normalsize,labelfont=sf,textfont=sf]{subfig}
\else
  \usepackage[caption=false,font=footnotesize]{subfig}
\fi

\usepackage{stfloats}
\usepackage{url}

\begin{document}
%
\title{Early Warnings of Cyber Threats in Online Discussions}

\author{\IEEEauthorblockN{Anna Sapienza*}\thanks{*A. Sapienza and A. Bessi contributed equally to this work.}
\IEEEauthorblockA{USC\\Information Sciences Institute\\
Marina del Rey, California, 90292}\\
\IEEEauthorblockN{Paulo Shakarian}
\IEEEauthorblockA{Arizona State University\\
Tempe, Arizona, 85281}
\and
\IEEEauthorblockN{Alessandro Bessi*}
\IEEEauthorblockA{USC\\Information Sciences Institute\\
Marina del Rey, California, 90292}\\
\IEEEauthorblockN{Kristina Lerman}
\IEEEauthorblockA{USC\\Information Sciences Institute\\
Marina del Rey, California, 90292}
\and
\IEEEauthorblockN{Saranya Damodaran}
\IEEEauthorblockA{Arizona State University\\
Tempe, Arizona, 85281}\\
\IEEEauthorblockN{Emilio Ferrara}
\IEEEauthorblockA{USC\\Information Sciences Institute\\
Marina del Rey, California, 90292}}


%


\maketitle

\begin{abstract}
We introduce a system for automatically generating warnings of imminent or current cyber-threats.
Our system leverages the communication of malicious actors on the darkweb, as well as activity of cyber security experts on social media platforms like Twitter. In a time period between September, 2016 and January, 2017, our method generated $661$ alerts of which about $84\%$
were relevant to current or imminent cyber-threats. In the paper, we first illustrate the rationale and workflow of our system, then we measure its performance.
Our analysis is enriched by two case studies: the first shows how the method could predict DDoS attacks, and how it would have allowed organizations to prepare for the Mirai attacks that caused widespread disruption in October 2016. Second, we discuss the method's timely identification of various instances of data breaches.
\end{abstract}


%
\IEEEpeerreviewmaketitle

\section{Introduction}
On October 21, 2016, hundreds of popular Websites, including Twitter, Netflix, and Paypal, became unreachable due to a massive cyber-attack directed against the infrastructure of the Internet. Attackers exploited a vulnerability in the software used by Internet of Things (IoT) devices that enabled them to commandeer a vast number of such devices. Attackers, then, used this so called botnet to unleash a massive Distributed Denial of Service (DDoS) attack against Dyn, a popular domain name system (DNS)   services provider. The overwhelmed Dyn DNS servers stopped routing Internet traffic effectively, resulting in widespread disruption of the involved Websites and therefore massive economic damage \footnote{https://www.theatlantic.com/technology/archive/2016/10/a-lot/505025/}. Despite the suddenness of the attack, there were signals indicating the availability of the IoT botnet in question as well as evidence of threat actors willing to employ this platform prior to the attack. This could have given decision makers key warnings to prepare for the imminent exploit.

To conduct a cyber-attack, malicious actors typically have to 1) identify vulnerabilities, 2) acquire the necessary tools and tradecraft to successfully exploit them, 3) choose a target and recruit participants, 4) create or purchase the infrastructure needed, and 5) plan and execute the attack. Other actors---system administrators, security analysts, and even victims---may discuss vulnerabilities or coordinate a response to attacks. These activities are often conducted online through social media, (open and dark) Web forums, and professional blogs, leaving digital traces behind. Collectively, these digital traces provide valuable insights into evolving cyber-threats and can signal a pending or developing attack well before malicious activity is noted on a target system. For example, exploits are discussed on Twitter before they are publicly disclosed~\cite{sabottke2015vulnerability} and on darkweb forums even before they are discussed on social media~\cite{nunes2016darknet}.

Here, we introduce a lightweight framework that leverages online social media sensors such as Twitter and darkweb forums, to generate alerts that function as early warnings of cyber-threats. The system monitors social media feeds of a number of prominent security researchers, analysts, and whitehat hackers, scanning for posts (tweets) related to exploits, vulnerabilities, and other relevant cyber-security topics.
Afterwards, it applies text mining techniques to identify important terms and remove irrelevant ones. Then, the system verifies whether the terms that were identified during the filtering stage have ever been used in darkweb hacking forums, and eventually reports the volume of mentions as well as the content of posts. Such information might be extremely valuable, since mentions that have been found by the algorithm might point to links to stolen credentials as well as threads where a novel vulnerability is discussed along with source codes aiming at exploiting it. Our framework relies on a database, updated daily, of posts published on nearly $200$ darkweb and deepweb hacking forums and marketplaces~\cite{mittal2016cybertwitter,DBLP:conf/isi/NunesDGMMPRSTS16,robertson2017darkweb}. Finally, the system generates warnings for the newly discovered terms, along with their frequency of appearance on social media and darkweb, the contents of possible mentions found in darkweb and deepweb, and a collection of words providing semantic context for facilitating situational awareness and interpretation of the warning. The algorithm design allows for generation of additional warnings over the same time period. This choice is due to the willingness to keep track of the attention around the possible cyber-threat, and in particular to monitor the evolution of darkweb activities related to the discovered terms.

To test the precision of the method we collected warnings generated between September 1, 2016 and January 31, 2017. Then, we asked to $5$ human annotators with 
knowledge of cyber-security to independently annotate each warning as a \emph{legitimate cyber-threat} or a \emph{false alarm} (see Section \ref{sec:performance} for additional details). We found that about $84\%$ of the total generated warnings were related to imminent or current cyber-threats, whereas the vast majority of false alarms were due to foreign languages terms (e.g. German) or terms related to cyber-security yet too generic to be considered as imminent or current threats.
In future versions of our algorithm, the number of false alarms could be 
reduced by adding dictionaries and updating the existing ones. The average time required to complete its entire workflow---from data retrieval to warning generation---is only $0.6$ seconds.  This makes our system amenable to real-time monitoring of multiple online platforms for imminent cyber-threats. Indeed, our system currently runs continuously in a cloud environment: all data are stored on Amazon EC2 and retrieved via Elastic Search, an open source distributed search engine that provides a powerful, scalable, and fast infrastructure for data retrieval. The endpoint of the system is designed to deliver early warnings to analysts and enable decision making allowing for preventive measures to face incoming attacks.

The rest of the paper is organized as follows. Section \ref{sec:algorithm} illustrates and explains the algorithm, from data retrieval to early-warnings generation. Section \ref{sec:performance} is dedicated to the evaluation of its performance and discovery accuracy. In Section \ref{sec:scenario}, we consider two different case studies, showing the use of the method to predict distributed denial of service attacks, specifically Mirai, and to detect instances of data breaches. Finally, in Section \ref{sec:discussion} we discuss future improvements as well as promising research directions.

\section{Method}
\label{sec:algorithm}
\begin{figure}[tbh]
\includegraphics[width = \columnwidth]{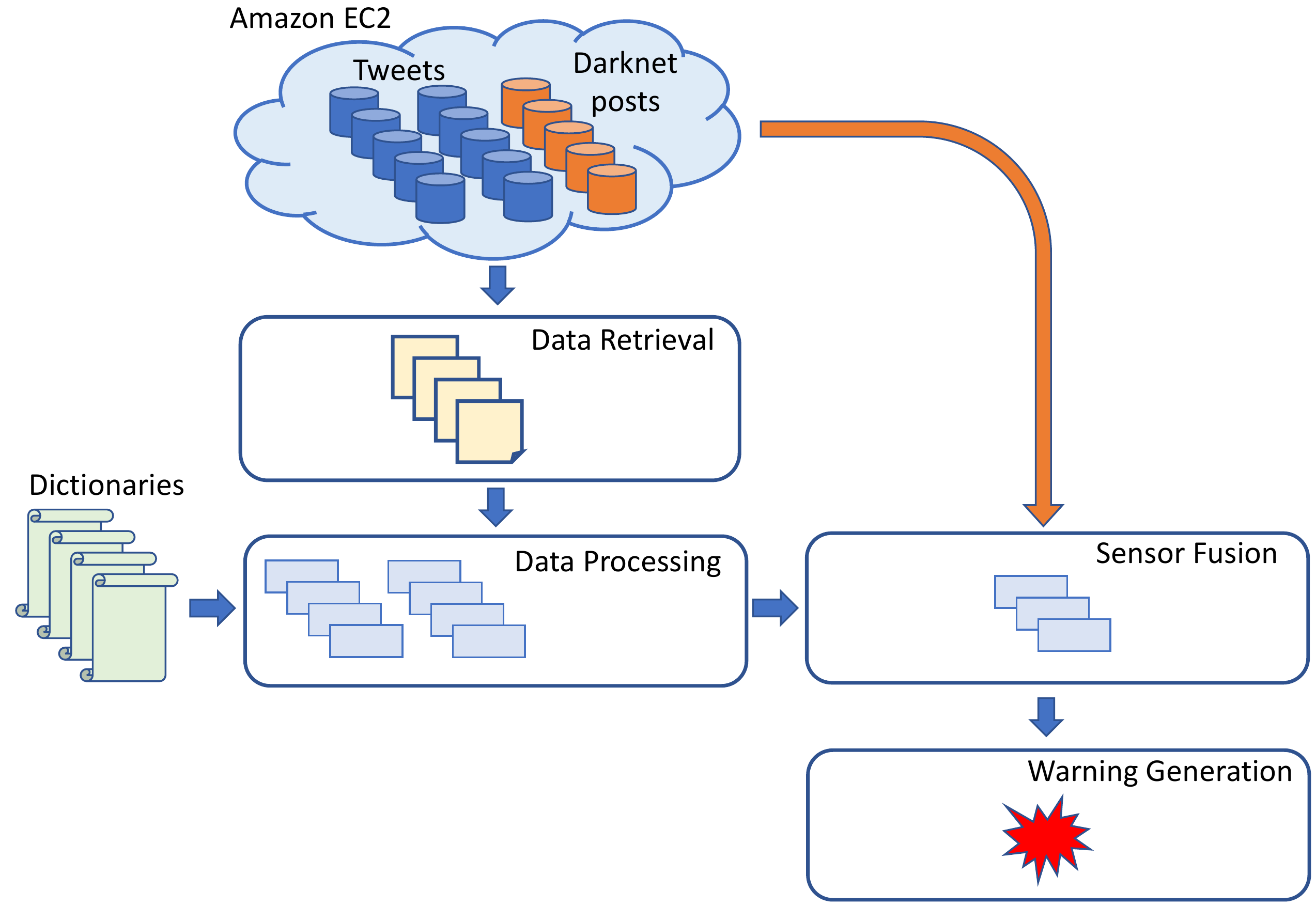}
\caption{\textbf{Workflow of the method.}}
\label{fig:workflow}
\end{figure}

In this section we provide a detailed explanation of our method, illustrating its workflow, from data retrieval to warning generation passing through data processing and sensor fusion. To facilitate the comprehension of our framework, we provide some practical examples throughout the section. Moreover, Figure \ref{fig:workflow} provides a graphical illustration of the system's workflow.

\subsection{Data Retrieval}
We compiled a list of well known and reliable Twitter experts who post frequently on issues related to cyber-security. This manually curated list includes $69$ international researchers and security analysts associated with security firms, as well as widely-followed whitehat hackers. The list can be arbitrarily extended by including other experts with similar degree of activity and trustworthiness.
We hourly collect all tweets posted in the previous $60$ minutes by these experts. The tweets are collected in real-time by means of Twitter API, stored in Amazon EC2, and retrieved using Elastic Search, an open source search engine based on Apache Lucene that provides a distributed, multitenant-capable full-text search with a schema-free JSON documents.

\paragraph*{Darkweb and deepweb crawling infrastructure.}
We briefly recap the infrastructure for crawling the darkweb and deepweb originally introduced in~\cite{robertson2017darkweb,DBLP:conf/isi/NunesDGMMPRSTS16}.  In this context, \emph{darkweb} refers to sites accessed through anonymization protocols such as Tor and i2p, while \emph{deepweb} refers to non-indexed sites on the open Internet~\cite{DBLP:books/sp/16/ShakarianGS16}.  Our crawling infrastructure handles sites of both types.  The framework consists of an infrastructure that enables lightweight crawlers and parsers that are focused on specific sites.  At the time of this writing, we have created crawlers and parsers for a manually-compiled list of nearly 200 sites relating to malicious hacking and/or online financial fraud, including fishing, spear-fishing, ransomware, credit card frauds, etc.  While our framework enables common crawling and parsing tasks based on aspects such as protocol and site structure, it remains imperative to use focused crawlers due to the variety of sites.  The use of customized parsers allow information from a variety of websites to be stored in a unified RDBMS schema which simplifies data cleaning~\cite{DBLP:conf/isi/NunesDGMMPRSTS16}, tagging~\cite{DBLP:conf/isi/MarinDS16}, and other analysis~\cite{DBLP:conf/aaai/RobertsonPSTS16}. These steps help ensure that the obtained data remains relevant to cyber-security use cases: indeed, many darkweb and deepweb forums and marketplaces also involve other illicit activities such as drug markets and the sale of stolen goods. To interface with the remainder of our method, we use a REST-based API, that allows to access posts' content and metadata such as publication date, authors' usernames, authors' reputations, etc.

\subsection{Data Processing}
After every iteration of the data collection, the tweets that have been retrieved are pre-processed using traditional text mining techniques: each tweet is lowercased, and numbers, symbols, Twitter handles and URLs are removed. At the end of this pre-processing stage, text is tokenized and all the tweets collected are aggregated and reduced to a unique list of terms.

For example, on September 5th 2016, between 8am and 9am GMT, a tweet that reads "\emph{My interview to $@$MalwareMustDie for $@$SecurityAffairs on a new Botnet targeting $\#$IoT. Details on $\#$Mirai trojan $\#$Linux }" has been retweeted seven times by the cyber-security experts that our system is monitoring. These seven identical tweets have been reduced to the following list of terms: \emph{my, interview, to, for, on, a, new, botnet, targeting, iot, details, mirai, trojan, linux}.

Each execution of the data pre-processing stage results in a very long list of terms, many of which are not related to cyber-threats. To address this issue, our framework automatically excludes any discovered term that appears in any of the following dictionaries:

\begin{itemize}
\item \texttt{english dictionary:} $235,892$ common English words---e.g. \emph{interview, new, details, ...};
\item \texttt{stopwords dictionary:} $2,390$ stopwords (for English, German, Italian, French, etc.)---e.g. \emph{to, on, a, for, ...};
\item \texttt{technical dictionary:} $57,459$ technical and context-specific terms that have been used from January, 2013 to August, 2016 by cyber-security experts the algorithm is monitoring---e.g. \emph{hacker, domain, dns, ...}; we manually created and curated this dictionary by analyzing the past six and a half years of activity of such experts on Twitter;
\item \texttt{threat dictionary:} $25$ general terms indicating known types of cyber-threats---e.g. \emph{ddos, phishing, databreach, botnet, etc}; we curated this list manually;
\item \texttt{italian dictionary:} $129,121$ common Italian words---e.g. \emph{intervista, attacco, spazio, ...}---needed since sometimes a few of the cyber-security experts tweet in Italian; other non-English dictionaries may be also used depending on the set of experts that the system leverages.
\end{itemize}

The four-stage filtering process allows the system to generate warnings only for terms that are likely to be related to new cyber-threats. Using \texttt{english dictionary}, \texttt{stopwords dictionary}, \texttt{italian dictionary}, we filter out common words that are unlikely to be related to cyber-threats; whereas by means of the \texttt{technical dictionary} we remove several context-specific words that have been used in the past by the cyber-security experts that we are monitoring.  Finally, those words that do not co-occur with terms contained in the \texttt{threat dictionary} are removed. This last stage of the filtering process is necessary to avoid the generation of warnings for words that because of their novelty are likely to pass through the first three filtering stages---e.g. words related to socio-political or novel trending topics discussed on Twitter.

Continuing the illustrative example, after the filtering process, the original list of terms obtained by tokenizing tweets has been reduced to a list containing only a single word: \emph{mirai}. The four-stage filtering process was able to filter out common words (\emph{interview, new, details}), stopwords (\emph{my, to, for, on, a}) and several context-specific words (\emph{targeting, botnet, linux, trojan, iot}) included in the dictionaries. Then, since the unique word left, \emph{mirai}, co-occurs with terms related to cyber-threats included in the \texttt{threat dictionary}, the algorithm proceeds to the third stage: sensor fusion.

\subsection{Sensor Fusion}
After raw data have been processed and irrelevant terms filtered out, the algorithm verifies whether the newly discovered terms have ever been used in a set of darkweb hacking forums. In fact, our algorithm relies on an up-to-date collection of all the posts that have been published in nearly $200$ darkweb hacking forums as well as social media. These posts are stored in Amazon EC2 and retrieved using Elastic Search. Continuing our example, in this stage the algorithm makes a Elastic Search query to look for occurrences of the word \emph{mirai} throughout all the posts that have been published on deepweb and darkweb forums. Eventually, the algorithm reports volume and content of posts mentioning the discovered term, often pointing either to links to stolen credentials or threads where a novel vulnerability is discussed.

\subsection{Warnings Generation}
In the final stage, the algorithm generates alerts providing some information related to it. 
Eventually, for each alert 
that passed all the four filtering stages and that occurred more than once in the observed period, the algorithm generates a warning reporting:

\begin{itemize}
\item The term that has been discovered and that is likely to be related to current or imminent cyber-threats;

\item The number of times that the discovered term has been mentioned on Twitter during the previous $60$ minutes;

\item The number of times that the discovered term has been mentioned on deepweb and darkweb hacking forums and marketplaces so far;

\item The content of posts where the discovered term has been mentioned on deepweb and darkweb hacking forums so far;

\item A collection of terms included in the \texttt{threat dictionary} that co-occurred in the tweets mentioning the discovered term. Such a collection of words provides semantic context for situational awareness that facilitates the interpretation of the generated warning.
\end{itemize}

Continuing our example, the warning that has been generated after data processing and sensor fusion is:
\begin{itemize}
\item time: current time;
\item threat: \emph{mirai};
\item frequency on Twitter: $7$;
\item frequency on Darknet/Deepweb: $0$;
\item posts on Darknet/Deepweb: empty;
\item context: \emph{botnet, linux, iot, trojan}.
\end{itemize}

\section{Precision and Performance Evaluation}
\label{sec:performance}

\begin{figure*}[tbh]
\includegraphics[width =  \textwidth]{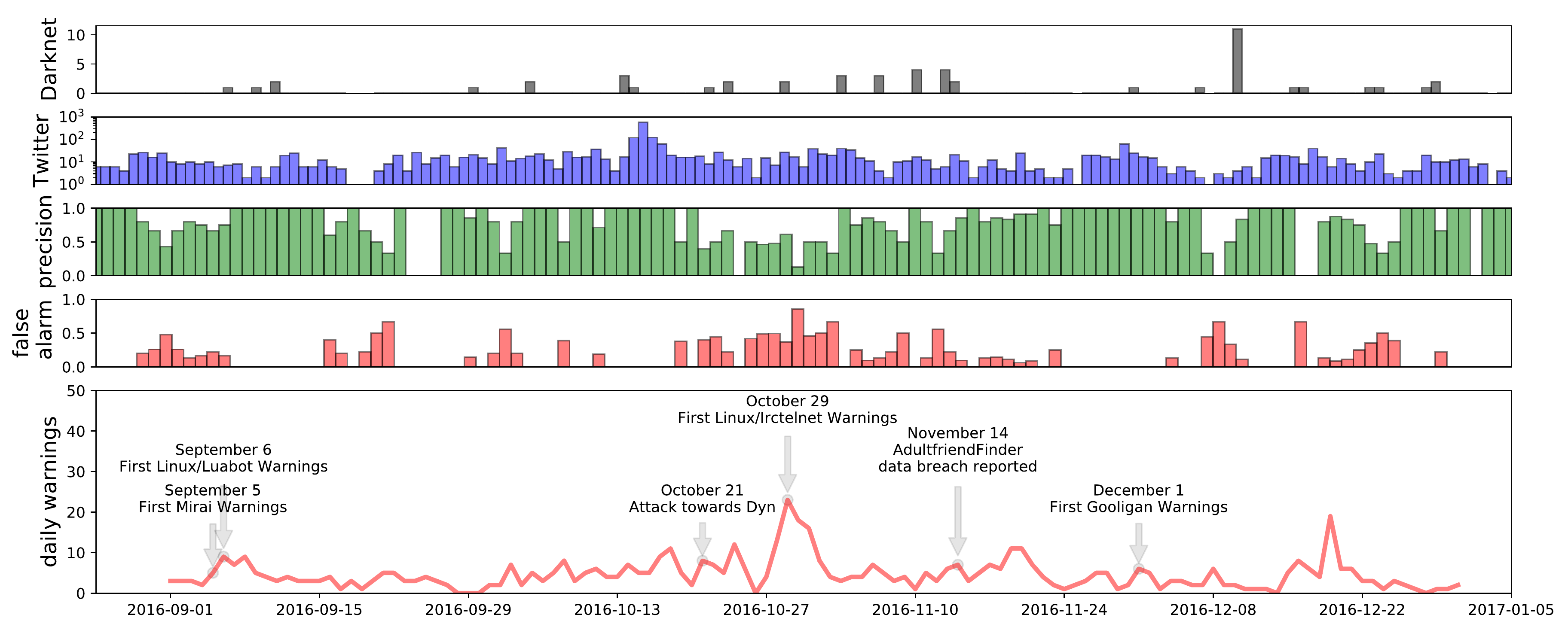}
\caption{\textbf{Number of warnings generated on a daily basis, along with their precision, the average fraction of experts agreeing on false alarms, and the amount of related tweets and darknet posts. Main events related to different types of attack are also annotated on the timeline.}}
\label{fig:precision}
\end{figure*}

To evaluate the accuracy of our framework, we ran it from September 1, 2016 to January 31, 2017 and collected all the warnings that have been generated during that period. Then, we asked to $5$ annotators with extensive knowledge in cyber-security to independently annotate each warning as \emph{legitimate cyber-threat} or \emph{false alarm}. Annotators were instructed to mark a warning as \emph{legitimate cyber-threat} if and only if the discovered word was strongly related to imminent or current cyber-threats. The annotators were further instructed about the possibility of using Google search to try identify attacks that occurred before, during, or after the warning generation time involving the specific threat. This enhanced the annotator's ability to determine whether the warning generated by our system was correlated to an actual attack or just a spurious coincidence (e.g., an old news regarding some old threat was retweeted and became contemporary even though no current or future threat occurred).

Over this period, the algorithm generated $661$ alerts, of which about $84\%$ have been considered \emph{legitimate cyber-threats}, from data breaches to novel vulnerabilities, by the majority (i.e. at least $3$) of the annotators. Table \ref{tab:annotators} reports precision according to each annotator.

\begin{table}
\centering
  \begin{tabular}{ | c | c | c | c |}
  \hline
  \textbf{annotator} & \textbf{threats} & \textbf{false alarm} & \textbf{precision (\%)} \\

  1 & 540 & 121 & 81.69\\
  2 & 532 & 129 & 80.48\\
  3 & 565 & 96 & 85.48\\
  4 & 534 & 127 & 80.79 \\
  5 & 578 & 83 & 87.44 \\

    \hline

    \hline

  \end{tabular}
\caption{\textbf{Algorithm's precision according to different annotators.}}
\label{tab:annotators}
\end{table}

Warnings annotated as false alarms were mostly represented by terms too generic to be considered legitimate cyber-threats or foreign language terms. Notice that the performance of our algorithm could be improved using additional dictionaries (e.g., other foreign languages) or adding new words to existent dictionaries.

Figure \ref{fig:precision} shows the number of warnings generated along with the average fraction of experts agreeing on false alarms, and the number of related posts occurring on Twitter and on the Darknet/Deepweb. For each day, we also show the achieved discover precision. Table \ref{tab:threats} shows some popular cyber-threats for which the algorithm generated several warnings. In particular, the algorithm found malware and randomware softwares, such as \textit{gooligan} and \textit{luabot}, both associated with disruptive attacks occurred during our observation period; vulnerabilities such as \textit{Mirai}; and data breaches such as \textit{AdultFriendFinder}, and \textit{BrazzersForum}, which we will discuss in the following section.

Concerning the execution performance, the average time required to complete the entire workflow---from data retrieval to warning generation---is $0.6$ seconds. Such a fast execution time is guaranteed by streamlining its entire workflow by means of Elastic Search, a distributed search engine that provides a powerful, scalable, and fast infrastructure for data retrieval.

\begin{figure*}[!h]
\centering
\includegraphics[width = 0.8\textwidth]{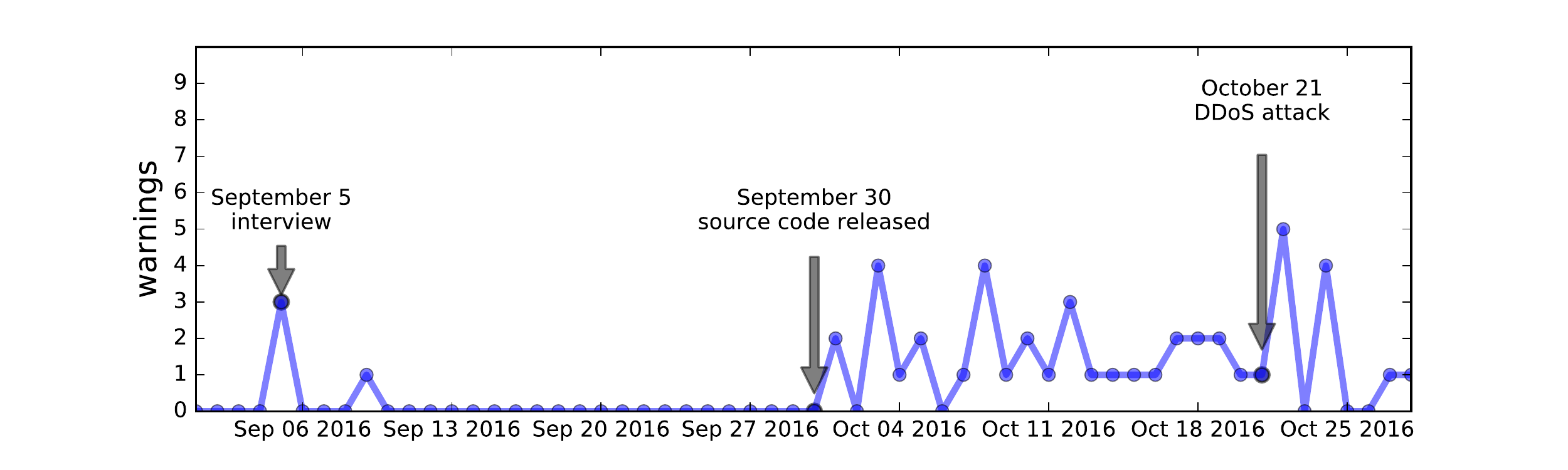}
\caption{\textbf{Timeline of Mirai warnings.}}
\label{fig:timeline_mirai}
\end{figure*}

\begin{table*}
\centering
  \begin{tabular}{ | c | c | c | c | c |}
  \hline
    \textbf{discovered} & \textbf{type of} & \textbf{number of} & \textbf{mentions on} & \textbf{mentions on} \\
  \textbf{word} & \textbf{threat} & \textbf{warnings} & \textbf{twitter} & \textbf{darkweb} \\
    \hline
    mirai & vulnerability & 94 & 537 & 85 \\
    teamxrat & ransomware & 13 & 30 & 0 \\
    luabot & trojan & 12 & 27 & 0 \\
    cryptoluck & ransomware & 12 & 29& 0 \\
    clixsense & data breach & 12 & 26 & 5 \\
    gooligan & malware & 9 & 26& 0 \\
    usbee & malware & 9 & 18 & 0 \\
    adultfriendfinder & data breach & 9 & 23& 0 \\
    starhub & data breach & 9 & 82 & 0 \\
    badepilogue & malware & 8 & 21 & 0 \\
    evony & data breach &8 & 25 & 0 \\
    \hline

  \end{tabular}
\caption{\textbf{Top cyber-threats anticipated by the method.}}
\label{tab:threats}

\end{table*}

\section{Scenario Analysis}
\label{sec:scenario}
In this section we provide evidences of the effectiveness of our method. In particular, we analyze two different scenarios, showing how our algorithm is able to identify imminent and current cyber-threats:

\begin{itemize}
\item \textbf{Vulnerabilities}: we will discuss how the method can be effectively used to timely predicting vulnerabilities that can be exploited in association with future cyber attacks. We will focus on the most prominent event during our analysis period (September 1, to January 31, 2017), namely Mirai and the disruption of the Dyn DNS operations occurred in October 2016;

\item \textbf{Data breaches}:  we will demonstrate how the method timely identified the availability of leaked data on darkweb/deepweb, as a consequence of data breaches due to cyber attacks; we will cover two instances of such data breaches, namely the publication on the darkweb of sensitive data on users of two platforms, AdultFriendFinder, and BrazzersForum.
\end{itemize}

\subsection{Vulnerabilities}

\paragraph{Mirai.}
In the introduction of this paper we mentioned the attack perpetrated on October 21, 2016 towards Dyn, a popular domain name system services provider. In this section we show how our algorithm generated warnings that could have allowed the  organization to anticipate and prepare for the attack.

The first warnings related to the vulnerability that eventually led to the aforementioned cyber attack were generated on September 5, 2016 between 7am and 9am GMT. A tweet linking to a blog post describing a vulnerability in the operating system of Internet of Things (IoT) devices was retweeted several times by the monitored cyber-security experts. Our algorithm generated two warnings for the term \emph{mirai} on that morning. The warnings generated by our system showed that the term \emph{mirai} was mentioned $14$ times between 7am and 9am GMT of September 5, 2016 along with terms related to cyber-threats such as \emph{malware, botnet, linux, iot, trojan}.

Mirai is a malware that turns computer systems running Linux---primarily IoT devices such as remote camera and home routers---into remotely controlled \emph{bots}, that can be used as part of a botnet in large-scale network attacks. Devices infected by Mirai continuously scan and identify further vulnerable IoT devices using a table of more than $60$ common factory default usernames and passwords; once a vulnerable device is identified, Mirai logs into it and infects it with a copy of its malware. Infected devices will continue to function normally, except for an increased use of bandwidth, associated to the scanning activity, and/or the execution of distributed denial of service attacks (DDoS). There are hundreds of thousands of IoT devices that use default settings, making them vulnerable to infection. Once infected, the device will monitor a command and control server which indicates the target of an attack. The use of a large number of IoT devices allows to bypass some anti-DDoS software which monitors the IP address of incoming requests and filters or sets up a block if it identifies an abnormal traffic pattern---e.g. if too many requests come from a particular IP address. Moreover, a large number of different infected device can provide more bandwidth than the perpetrator can assemble alone, as well as reduce the chance of being traced.

On October 1, the algorithm generated further warnings for the term \emph{mirai} and found that \emph{mirai} was mentioned several times on darkweb forums as well. Indeed, on September 30, the source code of a software kit exploiting this vulnerability was released on one of the darkweb forums that the method constantly monitors.

In the following days, several warnings were generated for the term \emph{mirai}, always emphasizing its co-occurrence with terms such as \emph{botnet} and \emph{iot}. In fact, the news of source code release quickly spread online, through blog posts and news stories, causing significant online chatter on Twitter and hacker forums that was promptly detected by our system. Figure \ref{fig:timeline_mirai} shows the timeline of warnings generated for the word \emph{mirai}.

Summarizing, the algorithm was able to generate warnings for a vulnerability with a lead time of $49$ days before hackers managed to exploit it to take over thousands of IoT devices and launch large scale distributed attacks. Moreover, since October 1---almost three weeks before the main attacks of October 21, 2016---the warnings were advising that the term \emph{mirai} was mentioned on darkweb/deepweb forums where users were sharing knowledge and source codes to exploit such a vulnerability, thus indicating that the threat was real and imminent.

This scenario illustrates the inherent predictability of such type of cyber attacks: the preparation of these events often occurs in plain sight, discussed on online platforms and publicly-accessible discussion forums. Our system leverages the human component that contributes to the attacks' preparation and leaves online digital traces behind.

\subsection{Data Breaches}
\paragraph{AdultFriendFinder.}
On November 14, 2016 several news outlets reported that more than $300M$ accounts of the popular adult dating website AdultFriendFinder were exposed in a massive data breach.

\begin{figure*}[!h]
\includegraphics[width = \textwidth]{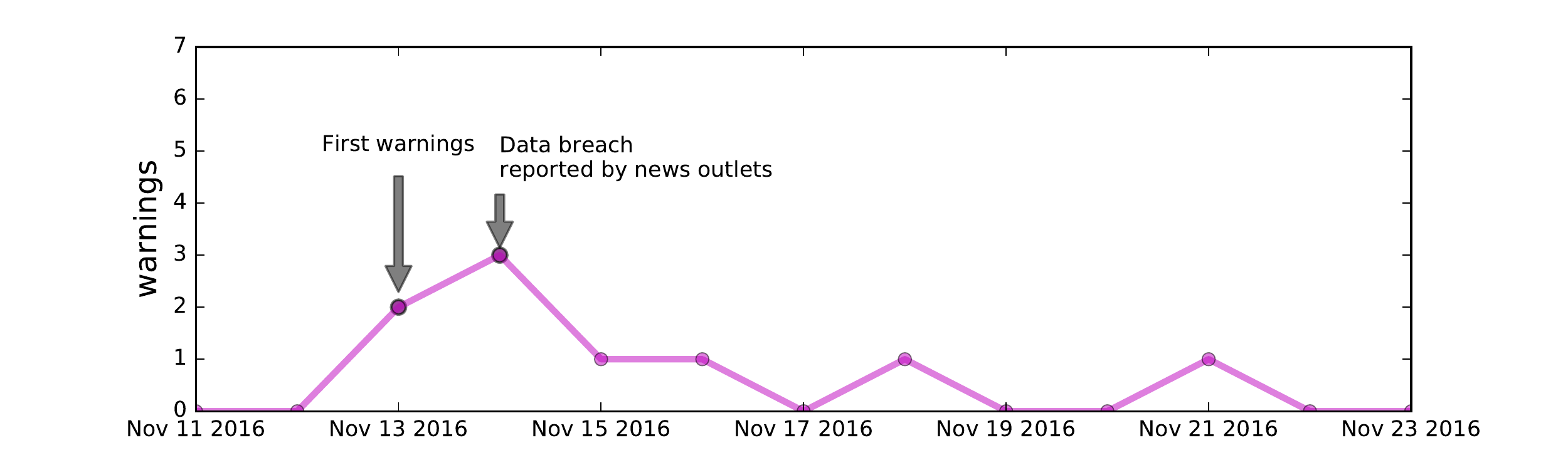}
\caption{\textbf{Timeline of AdultFriendFinder warnings.}}
\label{fig:timeline_adult}
\end{figure*}

The day before, on November 13, 2016 the algorithm started to issue several warnings related to the word \emph{adultfriendfinder} and associated with terms such as \emph{breach, accounts, databreach} (see Fig.~\ref{fig:timeline_adult}). Our warnings pointed to  $6$ mentions of \emph{adultfriendfinder} in darkweb and deepnet forums. Further inspection of these posts provided information about how to access to the leaked data, as well as links to marketplaces were the leaked data were sold.

\paragraph{BrazzersForum.}
On September 5, 2016 newspapers and magazines reported that nearly $800K$ accounts for popular adult website Brazzers were exposed in a data breach.

\begin{figure*}[!h]
\includegraphics[width =\textwidth]{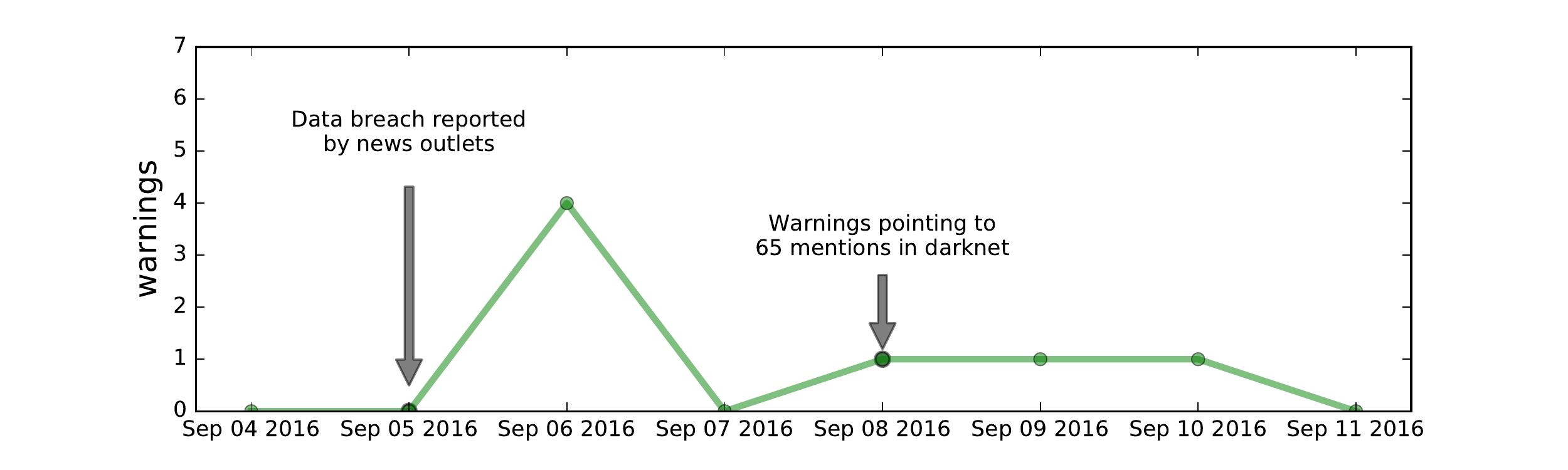}
\caption{\textbf{Timeline of BrazzersForum warnings.}}
\label{fig:timeline_brazzers}
\end{figure*}

On September 6, 2016 our algorithm generated $4$ warnings related to the word \emph{brazzersforum} and associated with terms such as \emph{accounts} and \emph{databreach} (see Fig.~\ref{fig:timeline_brazzers}). Initially, the algorithm did not find any mentions to \emph{brazzersforum} in deepnet and darkweb forums, but two days later, on September 8, the algorithm generated additional warnings indicating $65$ mentions to \emph{brazzersforum} in deepnet and darkweb forums. Such sources, again, provided various forms of access to these leaked data.

This case demonstrates that some events are more challenging to predict than others, since the system did not anticipate the data breach, differently from the AdultFriendFinder case. However, the method was yet capable of identifying sources that provide data availability, yielding actionable insights to analysts and decision makers to limit the damage of cyber-threats that have already occurred.

\section{Related Work}
The framework introduced in this paper leverages the communication of malicious actors on darkweb and deepnet hacking forums as well as the activity of cyber-security experts on Twitter.

Previous work analyzed hacker forums to detect threats that pose great risk to individuals, organizations, and governments. Research showed that the distribution of information among users is based on their skill level and reputation~\cite{holt2007subcultural,holt2012examining,jordan1998sociology}. Still, users disseminate tacit knowledge and share tools such as malware samples and source codes by simply attaching them to posted messages~\cite{fallmann2010covertly, benjamin2012securing}. Moreover, knowledge and methodology is disseminated  among hackers in the form of tutorials either written as text files or in the form of instructional videos~\cite{holt2012examining}. Many of these tutorials directly enable readers to launch criminal cyber attacks such as denial of services, SQL injections, cross-scripting attacks, and more \cite{benjamin2015exploring}. Researchers identified the presence of such hacker communities to be common across various geopolitical regions, including the US, China, Russia, the Middle-East, and other regions where information technologies are either ubiquitous or rapidly growing~\cite{motoyama2011analysis,benjamin2014time}.

Previous studies on online social media suggested that Twitter might serve as an important tool to predict future events in different domains, including earthquake detection~\cite{sakaki2010earthquake}, epidemiology~\cite{aramaki2011twitter,broniatowski2013national}, and the stock market~\cite{bollen2011twitter,bollen2011modeling}.
In the security domain, much work has been devoted to Twitter as object of manipulation and abuse~\cite{ferrara2015manipulation}, studying the effects of  spam accounts~\cite{wang2010don, benevenuto2010detecting}, social bots~\cite{ferrara2014rise, subrahmanian2016darpa}, and malicious campaigns~\cite{ratkiewicz2011detecting, thomas2012adapting, bessi2016social, varol2017online, ferrara2017disinformation}.
Sabottke \textit{et al.} published the first study to use  social media to identify cyber vulnerabilities~\cite{sabottke2015vulnerability}; we pushed this concept further and used security experts and their Twitter activity as  source for  cyber-threat anticipation.

\section{Discussion and Conclusions}
\label{sec:discussion}
In this paper we introduced a framework that generates warnings for imminent or current cyber-threats by leveraging unconventional, public data sensors such as Twitter and the Darknet. The performance evaluation reported in Section \ref{sec:performance} highlights that our system provides very high discovery precision, which could be easily improved by adding new dictionaries (e.g., foreign language dictionaries), enriching the existent one, and improving the threat dictionary to include new attacks terms. Indeed, the vast majority of false alarms reported by the annotators were due to foreign language terms and terms related to cyber-security yet too generic to be considered as real threats. The recall of our system may be improved by expanding the list of monitored experts, and adding further data sources.

In Section \ref{sec:scenario}, our analysis of the timeline of warnings pointed out how our algorithm could have allowed organizations to anticipate and prepare for attacks that caused wide spread disruptions in fall 2016. In particular, the system generated warnings for \emph{mirai}---the vulnerability that was exploited---49 days before hackers took over thousands of IoT devices and unleashed a massive DDoS attack that put the Internet infrastructure on its knees and disrupted the operations of hundreds of organizations, including banks and financial institutions, the entertainment industry, the US government, and various foreign countries. Furthermore, starting three weeks before the attack, the warnings were also identifying that the term \emph{mirai} was mentioned on darkweb hacker forums, in association with the availability of software toolkits to automate attack operations, stressing out that the threat was imminent even further.

However, we have noticed that for certain threats, such as the one related to \emph{BrazzersForums}, our algorithm is not able to produce warnings before the actual event occurs. To face this problem, we are now working on extending our algorithm to monitor multiple data sources at a time, e.g., Twitter, Darkweb forums, cyber security related Blogs, etc. This new extension will allow to generate warnings from several data sources and thus to detect cyber threat that might be not mentioned in Twitter or that might be mentioned in other sources earlier than in Twitter. As an example, by continuously monitoring the warnings generated by our method we were able to detect the two major attacks that occurred on May 12th, 2017 and on Jun 27th, 2017 respectively related to the ransomwares \emph{Wannacry} and \emph{Petya}. The method taking Twitter as a primary source produced the following warnings.
\paragraph{Wannacry} $24$ warnings related to the word  \emph{wannacry}, and $7$ to \emph{wannacrypt} were generated on May 12th. However, by monitoring other sources we have mentions in cyber security related Blogs of the words \emph{wannacry}, \emph{wannacrypt}, and \emph{wcry}, which first appear in Apr 18th and that our method could easily mark as warnings.
\paragraph{Petya}: $5$ and $3$ warnings related to the word \emph{petrwrap} (other name for \emph{petya}) were respectively generated back in March, 15th and March 21st. This signal disappears in Twitter until Jun 27th, the day of the cyber attack. The method was indeed able to generate early warnings related to this ransomware. However, the use of additional sources could help to continuously monitor the mentions of a discovered word. In the present, case we have seen mentions of \emph{petwrwrap} in the Darkweb forums both in March and April, 2017. 

Future work will be also devoted both to find a baseline model that could be used to compare the performance of our algorithm and to develop a method to automatically annotate the date in which a cyber threat becomes public. This improvement will enable to further evaluate the algorithm, by computing the average temporal length of the algorithm to discover new cyber threats in advance.

Finally, our method is still being improved with the aim to generate more detailed and informative warnings. Future versions of the algorithm will include a Natural Language Processing (NLP) stage aimed at extracting knowledge and insights from the Darkweb posts mentioning the discovered terms. In particular, we are developing NLP methods to recognize entities such as actors (hackers or groups), targets (organizations, specific sectors, etc.), source codes, etc.
We will consider extending our monitored keyword lists to layman terms (\textit{out-of-service, unavailable}, etc., instead of e.g., DDoS), monitor communities of open-source software developers (e.g., Android,  Linux, etc.) to timely identify new bugs and vulnerabilities as they become publicly known. 

We plan to leverage computational linguistic methods to investigate personality traits and socio-cultural traits of users mentioning the discovered words on darkweb forums: this will allow us to determine the credibility of a threat based on the expertise and the intents demonstrated by the actors associated to it.

\section{Acknowledgments}
The authors are supported by the Office of the Director of National Intelligence (ODNI) and the Intelligence Advanced Research Projects Activity (IARPA) via the Air Force Research Laboratory (AFRL) contract number FA8750-16-C-0112. The U.S. Government is authorized to reproduce and distribute reprints for Governmental purposes notwithstanding any copyright annotation thereon. Disclaimer: The views and conclusions contained herein are those of the authors and should not be interpreted as necessarily representing the official policies or endorsements, either expressed or implied, of ODNI, IARPA, AFRL, or the U.S. Government. 

The authors would like to thank Sindhu Kiranmai Ernala of Georgia Institute of Technology for her contribution to the paper. 

\balance
\bibliographystyle{IEEEtran}
\bibliography{biblio}

\end{document}